\documentclass[prd,aps,preprint,amsmath,nofootinbib,amssymb,eqsecnum]{revtex4-1}
\usepackage{verbatim,graphics,graphicx,color,slashed}
\usepackage{hyperref}

\begin{document}

\title{\Large Note on Spin-2 Particle Interpretation of the 750~GeV Diphoton Excess}

\author{Chao-Qiang~Geng$^{1,2,3}$\footnote{geng@phys.nthu.edu.tw}
and
Da~Huang$^{2}$\footnote{dahuang@phys.nthu.edu.tw}
}
  \affiliation{$^{1}$Chongqing University of Posts \& Telecommunications, Chongqing, 400065, China\\
  $^{2}$Department of Physics, National Tsing Hua University, Hsinchu, Taiwan\\
  $^{3}$Physics Division, National Center for Theoretical Sciences, Hsinchu, Taiwan
}

\date{\today}

\begin{abstract}
We explore the possibility to explain the 750 GeV diphoton excess recently measured by ATLAS and CMS collaborations in terms of a massive spin-2 particle. In particular, we consider the case in which the top-quark loops can give the similar amplitudes as the tree-level contributions to the diphoton and digluon channels. Such a scenario can be naturally realized in the generalized warped extra dimension models. As a result, the parameter space favored by the diphoton data implies that the top-quark loop contribution can compete to and even dominate over the tree-level amplitude in the spin-2 particle production via the gluon-gluon fusion process. Similar results are obtained for the case in which the spin-2 particle induces large invisible decay branching ratios, as indicated by the ATLAS data.

\end{abstract}

\maketitle

\section{Introduction}
Recently, a tantalizing hint for a resonance around 750~GeV has been observed by both ATLAS data with 3.2~${\rm fb}^{-1}$~\cite{ATLAS_diphoton} and CMS data with 2.6~${\rm fb}^{-1}$~\cite{CMS_diphoton}. This signal appears to be compatible with the LHC Run-1 data, which makes it a compelling candidate of long-waited new physics beyond the Standard Model (SM). It should be noted that, although the data from two experiments agree with each other very well, the best fit from ATLAS has a broad width of 45~GeV with a local significance of 3.9~$\sigma$, while CMS data favors a narrow resonance with a width of ${\cal O}(100~{\rm MeV})$ and a local significance of 2.6~$\sigma$. Furthermore, the null results in other searching channels such as $WW$, $ZZ$, $\ell \bar{\ell}$, $hh$ and dijets provide additional information on the possible couplings of the resonance to the SM particles.

Due to the Landau-Yang theorem~\cite{Landau:1948kw,Yang:1950rg}, the spin of the resonance 
accounting for 
the diphoton signal can only be 0 or 2. In the literature, there has already been a large amount of works~\cite{works} trying to interpret this diphoton resonance in various models beyond the SM, mostly focusing on the (pseudo-)scalar case. In the present paper, we consider an alternative explanation that the resonance is a massive spin-2 particle~\cite{Low:2015qep,Han:2015cty,Franceschini:2015kwy, Martini:2016ahj}, denoted as $G$, which can naturally arise as the Kaluza-Klein (KK) graviton in various contexts of the extra-dimensional models, such as ADD~\cite{ArkaniHamed:1998rs} and Randall-Sundrum (RS)~\cite{Randall:1999ee} types, with the motivation to solve the hierarchy problem~\cite{Antoniadis:1998ig,Han:1998sg}. It is interesting to note that the massive spin-2 particle can interact with the SM fields non-universally in some generalized versions of the RS model~\cite{WarpModel,Geng:2012hy}, with the interaction strengths depending on the localization of the SM fields in the bulk. In particular, we concentrate on a scenario in which the spin-2 particle's coupling with top quarks is much larger than those with the SM gauge fields, such as gluons and photons, so that the top-quark-loop contributions may play an important role in the spin-2 massive graviton production through the gluon-gluon fusion (ggF) and its decay to the diphoton, which is omitted in the previous discussions. It will be shown that such top-quark loop contributions can even dominate the production rate in some parameter space.

The paper is organized as follows. We begin with the short summary of the current experimental status of the diphoton signal and collect the limits of some relevant channels in Sec.~\ref{ExpSta}. In Sec.~\ref{Model}, we introduce our general framework for the interactions of the spin-2 massive graviton $G$ with SM fields, highlighting the origin of the hierarchy of couplings of $G$ between top quarks and the gauge fields. 
In Sec.~\ref{Pheno},
we perform  the computation of the production cross sections, widths, and branching ratios of the spin-2 particle, and then explore the parameter space which can explain the diphoton excess while satisfying other constraints from the LHC. We also consider the effects of invisible decay branching ratios to our analysis. Finally, a short summary is given in Sec.~\ref{Summary}.

\section{Diphoton Signal and Constraints From the LHC Run 1}\label{ExpSta}
Both ATLAS~\cite{ATLAS_diphoton} and CMS~\cite{CMS_diphoton} have reported the excess in the diphoton channel around 750~GeV
%, which could be explained by a new resonance. The diphoton 
with its production cross sections 
%in order to have such excess in ATLAS~\cite{ATLAS_diphoton} and CMS~\cite{CMS_diphoton} 
at 13 TeV are~\cite{Franceschini:2015kwy}
\begin{eqnarray}\label{ppXection}
\sigma (pp \to \gamma\gamma)_{\rm ATLAS} &\sim & (10\pm 3)~ {\rm fb}\,,\nonumber\\
\sigma (pp \to \gamma\gamma)_{\rm CMS} &\sim & (6 \pm 3)~ {\rm fb}\,,
\end{eqnarray}
respectively.

In Table~\ref{Tab_Constrait}, we collect the observed 95$\%$ upper limits for various channels from the 
LHC Run 1 data that will be applicable to our spin-2 particles.
\begin{table} [th]\caption{$95\%$ CL upper limits on the $\sigma \times {\cal BR}$ of a 750 GeV resonance decaying to various final states from the 
8 TeV LHC Run 1 data. The numbers in the last column give the LHC limits on the $\sigma \times {\cal BR}$ at 13 TeV obtained by rescaling from 8 TeV using the $gg$ parton luminosity.}
\begin{center}
\begin{tabular}{c||cc||c}
Final States    & Observed & Experiment & Rescaled \\ \hline\hline
$\gamma\gamma$  & 2.5 fb   & ATLAS~\cite{Aad:2015mna} & 11.7 fb  \\
$\gamma\gamma$ (narrow width) &  1.5 fb & CMS~\cite{Khachatryan:2015qba} & 7.0 fb  \\
$\gamma\gamma$ (large width)  &  2.0 fb & CMS~\cite{Khachatryan:2015qba} & 9.4 fb \\
$\gamma\gamma$ (spin-2)       &  1.8 fb & CMS~\cite{CMS:2015cwa} & 8.5 fb \\ \hline
$t\bar{t}$ (scalar)  &  960 fb  & ATLAS~\cite{Aad:2015fna}   & 4.5 pb    \\
$t\bar{t}$ (vector)  &  790 fb  & ATLAS~\cite{Aad:2015fna}   & 3.7 pb    \\
$t\bar{t}$ (narrow)  &  450 fb  & CMS~\cite{Khachatryan:2015sma} & 2.1 pb \\
$t\bar{t}$ (wide)    &  520 fb  & CMS~\cite{Khachatryan:2015sma} & 2.4 pb \\ \hline
$ZZ$ ($\ell\ell jj$) &   40 fb  & ATLAS~\cite{Aad:2014xka}    & 180 fb   \\
$ZZ$ ($\ell\ell jj$) &   50 fb  & CMS~\cite{Khachatryan:2014gha} & 230 fb \\
$ZZ$ (combined)      &   10 fb  & ATLAS~\cite{Aad:2015kna}    & 46 fb     \\ \hline
$Z\gamma$ ($\ell\ell \gamma$)  &   4.0 fb  & ATLAS~\cite{Aad:2014fha}  & 19 fb     \\ \hline
$jj$\footnote{The dijet limit is set on $\sigma \times {\cal BR} \times {\cal A}$ with ${\cal A}$ is the acceptance.}        & 14.0 pb  & ATLAS~\cite{Aad:2014aqa}  & 65  pb     \\
\end{tabular}
\label{Tab_Constrait}
\end{center}
\end{table}
It is noted that the diphoton signal rate of ${\cal O}(10~ {\rm fb})$ from the 13 TeV data is large, 
which implies that  there should be a large enhancement from 8 TeV to 13 TeV for the signal as compared with the background. 
Hence, it was pointed in Ref.~\cite{Han:2015cty} that the ggF is the preferred production channel than the diquark process. 
In the present paper, we focus on a massive spin-2 particle model in which $G$ is mostly produced by the ggF in order 
to take advantage of this fact. With this assumption, we also give the corresponding 13 TeV limits  rescaled  from the 8 TeV ones 
by the $gg$ parton luminosity ratio of 4.7~\cite{TWIKI} in the last column of Table~\ref{Tab_Constrait}. 
Although a strict study of the compatibility of a proposed model with the 8 TeV limits needs to simulate the signals and backgrounds, 
such a rescaling offers a quick and easy way to obtain the constraints on the model. 
Moreover, by comparison, it is seen that the required cross sections for the 750~GeV diphoton excess in Eq.~(\ref{ppXection}) 
are fully compatible with the 8~TeV data of both ATLAS and CMS.

As mentioned in the Introduction, the ATLAS data favors a wide resonance of width 45~GeV with the local significance of 3.9~$\sigma$, 
whereas the CMS best fit indicates a narrow resonance. However, as we shall see later, for a generic spin-2 particle with reasonable couplings, the total decay width from the SM channels is always about of only ${\cal O}(1~{\rm GeV})$. Therefore, in order to accommodate the ATLAS result, there should be other invisible decays for the massive spin-2 particle, possibly connected to  dark matter. In our following phenomenological discussions, we shall consider the cases with the narrow and wide resonances, respectively.

\section{General Framework of the Massive Graviton}\label{Model}
The spin-2 particle $G$ considered in the present paper originates from the first KK graviton in the generalized RS model~\cite{Randall:1999ee} in which some of the SM fields propagate in the extra fifth dimension, while others localized on either  UV or IR branes~\cite{WarpModel,Geng:2012hy}. The general interactions between this massive graviton with the SM particles can be parametrized as follows,
\begin{eqnarray}
{\cal L} = -\sum_i\frac{c_i}{\Lambda} G_{\mu\nu} T_i^{\mu\nu}\,,
\end{eqnarray}
where $T_i^{\mu\nu}$ and $c_i$ stand for the energy-momentum tensors of the SM particles and their coupling strengths to the spin-2 particle of $G_{\mu\nu}$, 
respectively, and $\Lambda$ is the typical interaction energy scale. Note that $c_i$ depend on the overlap among the wave functions between 
the KK graviton and the corresponding SM particles. Since the KK graviton is localized towards the IR brane, 
it has unsuppressed couplings to the fields on the IR brane or in the bulk, and greatly suppressed couplings to the ones sitting on the UV brane. 
Thus, a natural hierarchy among couplings can be obtained. For simplicity, we consider the following field configurations: only the right-handed top quark is put on the IR brane, and the SM hypercharge $U(1)_Y$ gauge field $B_\mu$ and the $SU(3)_C$ gluon $G_\mu$ propagate in the bulk, while other SM fields are localized on the UV brane, including the $SU(2)_L$ gauge and  Higgs bosons. Moreover, the couplings of $G$ to the constant zero mode of bulk gauge fields in the RS models suffer from an extra mild suppression by the extra-dimension volume factor $c_{gg}\sim c_1 \sim 1/\ln(M_P/M_{\rm IR})\sim 0.03$ with the IR brane scale $M_{\rm IR}$ of ${\cal O}({\rm TeV})$, which is similar to the order of the corresponding one-loop factor $\alpha/(4\pi)$ ($\alpha_s/(4\pi)$) for the
 bulk EW gauge fields and gluons. By taking this fact into account, $c_{tt}$ should be of ${\cal O}(1)$, while the couplings for the bulk gauge fields $B_\mu$ and $G_\mu$ should be rescaled to $\alpha c_1/(4\pi) $ and $\alpha_s c_{gg}/(4\pi)$. Therefore, the expected hierarchy of various KK graviton couplings is given by
\begin{eqnarray}\label{hierarchy}
c_{tt} > \alpha_s c_{gg}/(4\pi) \sim \alpha c_{1}/(4\pi) \gg c_{\rm others} .
\end{eqnarray}

After the spontaneous EW symmetry breaking, the relevant couplings of this massive KK graviton in the basis of the SM particle mass eigenstates become
\begin{eqnarray}\label{LagG}
{\cal L}_G &=& -\frac{1}{\Lambda} G_{\mu\nu} \Big[ \frac{\alpha c_{\gamma\gamma}}{4\pi} \left(\frac{1}{4} \eta^{\mu\nu} A^{\lambda\rho} A_{\lambda\rho} - A^{\mu\lambda}A^{\nu}_{\lambda}\right) +\frac{\alpha c_{Z\gamma}}{4\pi} \left( \frac{1}{4} \eta^{\mu\nu} A^{\lambda\rho} Z_{\lambda\rho} - A^{\mu\lambda}Z^{\nu}_{\lambda} \right) \nonumber\\
&& + \frac{\alpha c_{ZZ}}{4\pi} \left( \frac{1}{4} \eta^{\mu\nu} Z^{\lambda\rho} Z_{\lambda\rho} - Z^{\mu\lambda}Z^{\nu}_{\lambda} \right) + \frac{\alpha_s c_{gg}}{4\pi}  \left( \frac{1}{4} \eta^{\mu\nu} G^{a\, \lambda\rho} G^a_{\lambda\rho} - G^{a\, \mu\lambda}G^{a\, \nu}_{\lambda} \right) \nonumber\\
&& + c_{tt} \Big(\frac{i}{4} \bar{t}_R (\gamma^\mu \partial^\nu + \gamma^\nu \partial^\mu)t_R -\frac{i}{4} (\partial^\mu \bar{t}_R \gamma^\nu + \partial^\nu \bar{t}_R \gamma^\mu) t_R \nonumber\\
&& -i \eta^{\mu\nu} [\bar{t}_R \gamma^\rho \partial_\rho t_R - \frac{1}{2}\partial^\rho (\bar{t}_R\gamma_\rho t_R)]   \Big) \Big] \,,
\end{eqnarray}
where the couplings $c_{\gamma\gamma, Z\gamma, ZZ}$ are related to the original $c_1$ via
\begin{eqnarray}
&& c_{\gamma\gamma} = c_1 \cos^2\theta_W , ~ c_{Z\gamma} = -c_1 \sin 2\theta_W = -\sin 2\theta_W c_{\gamma\gamma}/\cos^2\theta_W, ~\nonumber\\
&& c_{ZZ} = c_1 \sin^2\theta_W = \tan^2\theta_W c_{\gamma\gamma} \,.
\end{eqnarray}
As a result, we can use $c_{gg}$, $c_{tt}$ and $c_{\gamma\gamma}$ to parametrize this model. Although it cannot explain the hierarchy problem in terms of the warped factor in the extra dimension, the present model is the simplest setup for the spin-2 particle to interpret the 750 GeV resonance with the emphasis on the important top-quark loop contributions in its production and decays.

One may wonder why we neglect spin-2 particle couplings to $W$-bosons in Eq.~(\ref{LagG}). Firstly, according to our bulk field configuration and the hierarchies of couplings  in Eq.~(\ref{hierarchy}), the direct coupling of the massive spin-2 particle $G$ to the $W^+ W^-$ pair is vanishingly small. Moreover, we only allow $G$ to couple the right-handed top quark via the $O(1)$ coupling $c_{tt}$, 
resulting in that the induced one-loop effective $GW^+W^-$ coupling would be suppressed by an
 additional factor $(m_t/m_G)^2 \sim 0.05$ compared with the corresponding $\gamma\gamma$ and $gg$ couplings. 
 As a result, we expect that the decay rate of this channel is too small to be observed.

\section{Phenomenological Explanation of Diphoton Excess}\label{Pheno}
\begin{figure}[th]
\includegraphics[scale=0.7]{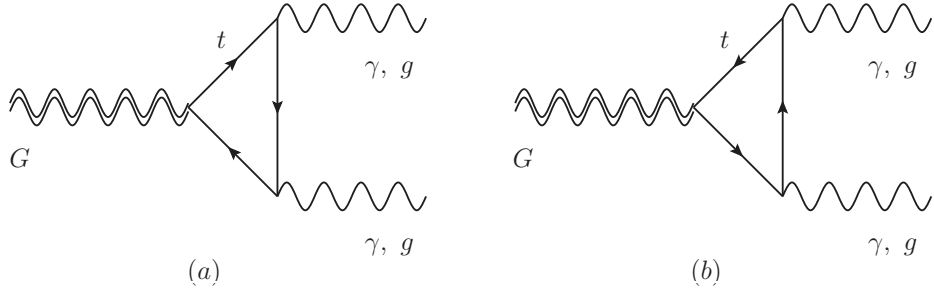}
\caption{Feynman Diagrams for top-quark loop contributions to the effective couplings of the spin-2 particle with gluons and photons. }
\label{FeynD}
\end{figure}

\begin{figure}[th]
\includegraphics[scale=0.7]{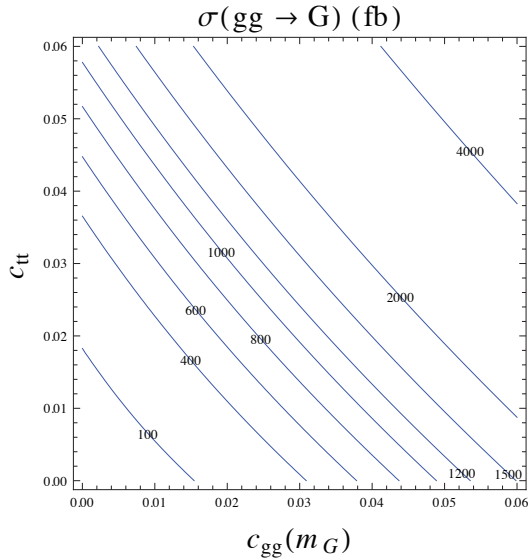}
\caption{The production cross section of the spin-2 particle $G$ as a function of the couplings $c_{gg}(m_G)$ and $c_{tt}$, where the number in each line denotes the corresponding value of the cross section in unit of fb.}
\label{ProdG}
\end{figure}

The production of the spin-2 particle $G$ is mostly via the ggF process in our scenario. Since the top-quark loop contribution to the amplitude of 
the $G$ production is expected to have the same order as the tree-level one, it is useful to define an effective coupling between $G$ and gluons:
\begin{eqnarray}\label{cgg}
c_{gg}^{\rm eff} = c_{gg}(m_G) + \frac{c_{tt}}{3} A_G\left(\frac{4m_t^2}{m_G^2}\right)\,,
\end{eqnarray}
where
\begin{eqnarray}
A_G(\tau) &=& -\frac{1}{12}\Big[ -\frac{9}{4} \tau (\tau + 2) [2{\rm tanh}^{-1}(\sqrt{1-\tau}) - i \pi]^2 \nonumber\\
&& + 3 (5\tau + 4) \sqrt{1-\tau}[2{\rm tanh}^{-1}(\sqrt{1-\tau})-i\pi] \nonumber\\
& &  -39 \tau + 12\ln \tau -35 - 12\ln4  \Big]\,,
\end{eqnarray}
is the top-quark loop function for the spin-2 particle, which is valid for $\tau < 1$ and obtained by calculating the Feynman diagrams in Fig.~\ref{FeynD}.
It follows that the massive graviton production cross section mainly depends on two couplings, the
tree-level gluon coupling $c_{gg}(m_G)$ renormalized at $m_G = 750$~GeV and the top-quark coupling $c_{tt}$. 
Consequently, the $G$ production cross section can be conveniently plotted in the $m_{gg}(m_G)$-$c_{tt}$ plane as in Fig.~\ref{ProdG} with $\Lambda = 1$~TeV. Note that in our calculation we have applied the RS model~\cite{RSmodel} with {\sf Feynrules}~\cite{Christensen:2008py, Degrande:2011ua} to the {\sf MadGraph}~\cite{Alwall:2011uj}.

Given the interactions in Eq.~(\ref{LagG}), the partial decay widths for various channels can be calculated. 
Here, we only present the analytical formula of the most relevant ones, such as $\bar{t} t$, $gg$, $\gamma\gamma$, $Z\gamma$ and $ZZ$,
% Since only the right-handed top-quark contribution, so the amplitude of each decay process should be smaller by a factor of 2, so the total width and production rate should be smaller by a factor of 4 for the top-quark loop contributions. And for $t\bar{t}$ decay channel, the decay width will be smaller by a factor of 2 due to the same reason.
given by
\begin{eqnarray}
\Gamma(G\to t\bar{t}) &=& \frac{N_c}{320\pi} \frac{c_{tt}^2 m_G^3}{\Lambda^2} \left(1-\frac{4m_t^2}{m^2_G}\right)^{3/2} \left(1+\frac{8}{3} \frac{m_t^2}{m_G^2}\right)\,, \label{ttRate} \\
\Gamma(G\to \gamma\gamma) &=& \frac{m_G^3}{80\pi\Lambda^2} \left|\frac{\alpha}{4\pi} c_{\gamma\gamma}^{\rm eff}  \right|^2\,, \\
\Gamma(G\to gg) &=& \frac{m_G^3}{10\pi\Lambda^2} \left|\frac{\alpha_s c_{gg}^{\rm eff}}{4\pi}\right|^2\,,\\
\Gamma(G\to Z\gamma) &=& \frac{m_G^3}{160\pi\Lambda^2} \left(\frac{\sin 2\theta_W}{\cos^2 \theta_W}\right)^2 \left(1-\frac{m_Z^2}{m_G^2}\right)^3 \left(1+ \frac{m_Z^2}{2m_G^2} + \frac{m_Z^4}{6m_G^4}\right)  \left|\frac{\alpha c_{\gamma\gamma}^{\rm eff}}{4\pi} \right|^2 \, \label{ZpRate} , \\
\Gamma(G\to ZZ) &=& \frac{m_G^3}{80\pi\Lambda^2} \tan^4\theta_W \left(1-\frac{4 m_Z^2}{m_G^2}\right)^{1/2} \left(1 - \frac{3 m_Z^2}{m_G^2} + \frac{6 m_Z^4}{m_G^4}\right) \left|\frac{\alpha c_{\gamma\gamma}^{\rm eff}}{4\pi} \right|^2 \,, \label{ZZRate}
\end{eqnarray}
with the effective coupling of $G$ with photons as
\begin{equation}
c_{\gamma\gamma}^{\rm eff} = c_{\gamma\gamma} (m_G) + \frac{2}{3}  Q_t^2 N_c c_{tt}  A_G \left(\frac{4 m_t^2}{m_G^2}\right)\,,
\end{equation}
and $c_{gg}^{\rm eff}$ as defined in Eq.~(\ref{cgg}),
where $N_c = 3$, $Q_t = 2/3$ and $m_t = 173.5$~GeV~\cite{PDG} denote the color, electric charge and mass of the top quark, respectively. Note that we 
have ignored the $Z$ boson mass in the expressions of the top-quark loop contributions for the $Z\gamma$ and $ZZ$ modes since its effect is greatly suppressed by the spin-2 particle mass $m_G = 750$~GeV. In Appendix, we present the details of the calculations of the top-quark loop amplitudes for the final states of $\gamma\gamma$, $Z\gamma$, $ZZ$ and $gg$, as shown in Fig.~\ref{FeynD}, with the emphasis of the issue of the renormalization. In the remaining part of this section, we assume that the spin-2 particle only decays through these visible channels so that a narrow width of ${\cal O}(1~\mbox{GeV})$ is predicted, while leaving the discussion of the case with a broad width due to its large invisible decay to the next section.
%  Moreover, due to the different power counting in the present model, the interaction in Eq.~(\ref{LagG})   the formulae for the decay widths of $Z\gamma$ and $ZZ$ modes are different from those in Ref.~\cite{Han:1998sg} due to the difference of the effective Lagrangian in Eq.

\subsection{Constraints From $t\bar{t}$ Channel}
\begin{figure}[th]
\includegraphics[scale=0.7]{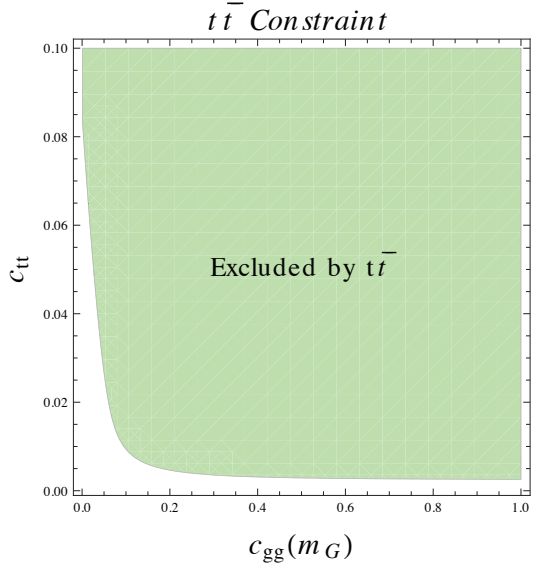}
\caption{The constraint from $gg\to G \to t \bar{t}$ in the $c_{gg}(m_G)$-$c_{tt}$ plane with $c_{\gamma\gamma} (m_G) = 0.1$ and $\Lambda = 1$~TeV,
where the green region is excluded by the CMS data~\cite{Khachatryan:2015sma}.}
\label{Const_tt}
\end{figure}

We begin with our discussion with the decay of the spin-2 particle into $t\bar{t}$ since it gives the most stringent constraints on the present scenario due to 
the non-suppressing couplings of $G$ with top quarks. From Eq.~(\ref{ttRate}), it is useful to present this constraint in the $c_{gg}(m_G)$-$c_{tt}$ plane 
as in Fig.~\ref{Const_tt} with $c_{\gamma\gamma} = 0.1$ and $\Lambda = 1$~TeV.
Note that the variation of  $c_{\gamma\gamma}$ from $10^{-4}$ to 1.0 does not change the plot much, since the branching ratios of the 
spin-2 particle would be always dominated by $t\bar{t}$ unless  $c_{tt}$ is chosen to be very small. From Fig.~\ref{Const_tt}, it is interesting that the $t\bar{t}$ upper bound naturally separates the viable parameter space into two regions: relatively large $c_{tt}$ but small renormalized tree-level coupling $c_{gg}(m_G) < 0.1$, and large $c_{gg}(m_G)$ but nearly vanishing $c_{tt} < 0.01$. Since the tree-level spin-2 particle couplings to photons and gluons are parametrized of the same order to the corresponding top-quark loop contributions, we expect that the above two parameter regions correspond to the cases with the top-quark loop effectively turned on or  off. Therefore, in the following we shall consider the two benchmark scenarios with $c_{tt} = 0$ and $c_{tt} \neq 0$. For simplicity,
 we fix the cutoff scale $\Lambda = 1$~TeV throughout our study.

\subsection{Parameter Space with $c_{tt} = 0$}
\begin{figure}[th]
\includegraphics[scale=0.7]{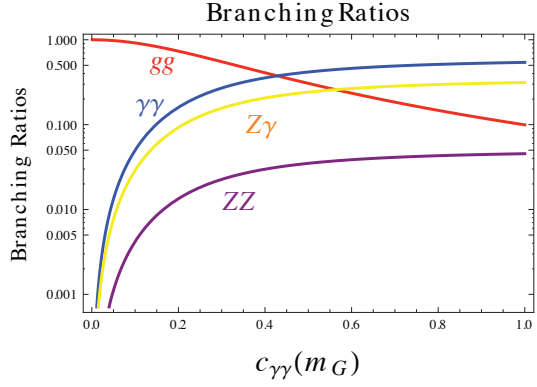}
\caption{Branching ratios of the spin-2 particle $G$ as  functions of $c_{\gamma\gamma}(m_G)$ which parametrize the tree-level contribution of $G$ to the diphoton decay, where  $c_{tt} = 0$, $c_{gg} (m_G) = 0.01$, and $\Lambda = 1$~TeV are used. }
\label{BR0}
\end{figure}

In this subsection, we shall explore the parameter space of the spin-2 particle with $c_{tt} = 0$ so that the top-quark loops to $\gamma\gamma$ and $gg$ do not contribute. In Fig.~\ref{BR0}, we show the typical branching ratios as functions of $c_{\gamma\gamma}(m_G)$ 
%in this parameter region 
with  $c_{tt} = 0$ and $c_{gg} (m_G) = 0.01$. It is seen that the dijet ($gg$) dominates the branching ratio unless $\gamma\gamma$ and $Z\gamma$ are greatly enhanced by the tree-level coupling $c_{\gamma\gamma} (m_G)$.

\begin{figure}[th]
\includegraphics[scale=0.7]{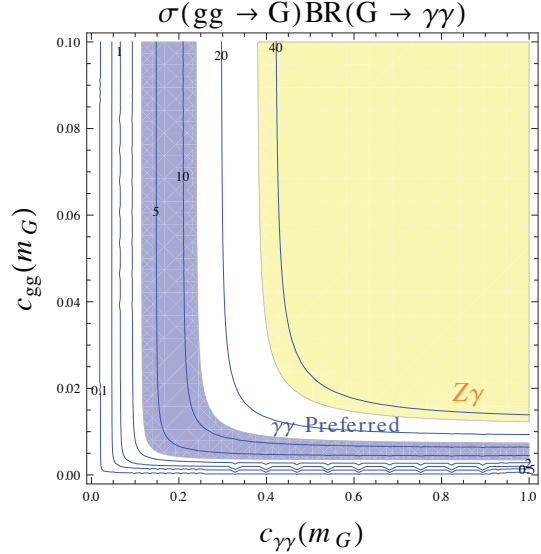}
\caption{Parameter space of the spin-2 particle favored by the $\gamma\gamma$ excess observed by ATLAS and CMS (Blue Region), together with the constraint from $Z\gamma$ (Yellow Region), where the numbers on the blue lines denote the corresponding $\gamma\gamma$ production cross sections due to 
the spin-2 particle decays. }
\label{pp0}
\end{figure}

From Fig.~\ref{pp0}, one can see that it is possible to obtain the measured $\gamma\gamma$ signal rate of $3\sim 13$ fb as implied by the ALTAS and CMS data. 
In addition, the most relevant limit is the $Z\gamma$ bound, which, however, does not constrain the $\gamma\gamma$ preferred region at all. Note that this scenario was already investigated in Refs.~\cite{Han:2015cty, Franceschini:2015kwy, Martini:2016ahj}, which obtained the similar conclusions.

\subsection{Parameter Space with $c_{tt} \neq 0$}
\begin{figure}[th]
\includegraphics[scale=0.7]{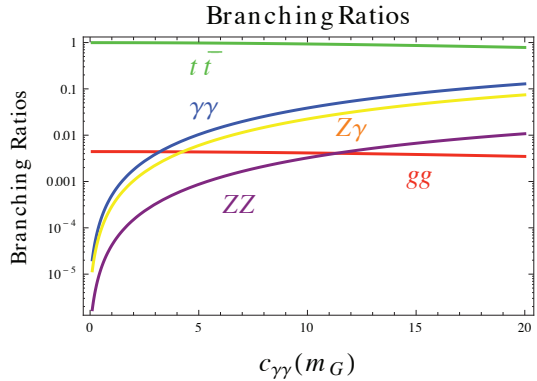}
\caption{Branching ratios of the spin-2 particle $G$ as  functions of $c_{\gamma\gamma}(m_G)$ which parametrize the tree-level contribution of $G$ to the diphoton decay, where  we have taken $c_{tt} = 0.04$, $c_{gg}(m_G)=0.03$, and $\Lambda = 1$~TeV. }
\label{BR1}
\end{figure}

We now turn to the case with $c_{tt} \neq 0$.  The decay branching ratios as  functions of $c_{\gamma\gamma}(m_G)$ for various channels are shown in Fig.~\ref{BR1} with the typical parameter choices of $c_{tt} = 0.04$ and $c_{gg} (m_G) = 0.3$, which are allowed by the $t\bar{t}$ constraint presented 
in Fig.~\ref{Const_tt}.
As expected, the width is dominated by $t\bar{t}$ modes due to unsuppressed couplings to $G$, but $\gamma\gamma$, $Z\gamma$ and $ZZ$ can become 
sizable as $c_{\gamma\gamma} (m_G)$ increases to be large enough.

\begin{figure}[th]
\includegraphics[scale=0.7]{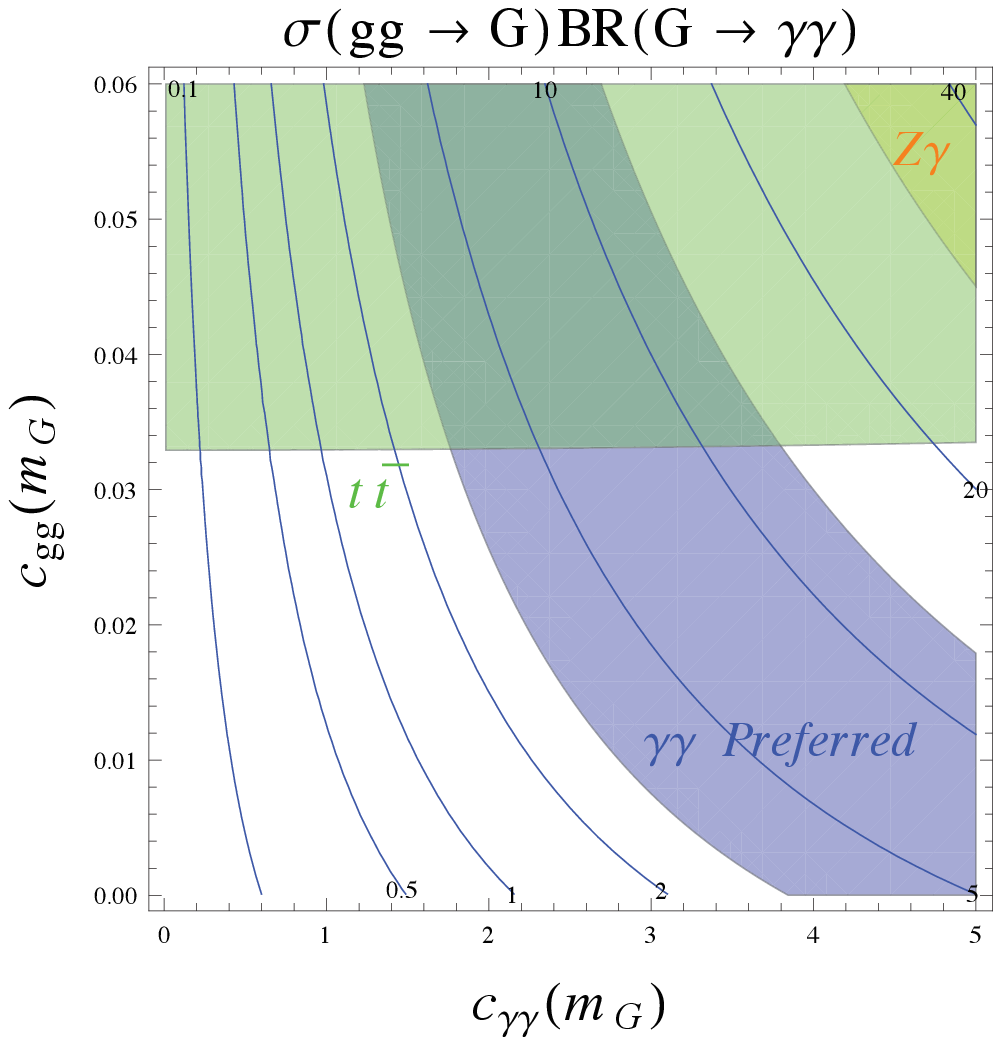}
\includegraphics[scale=0.7]{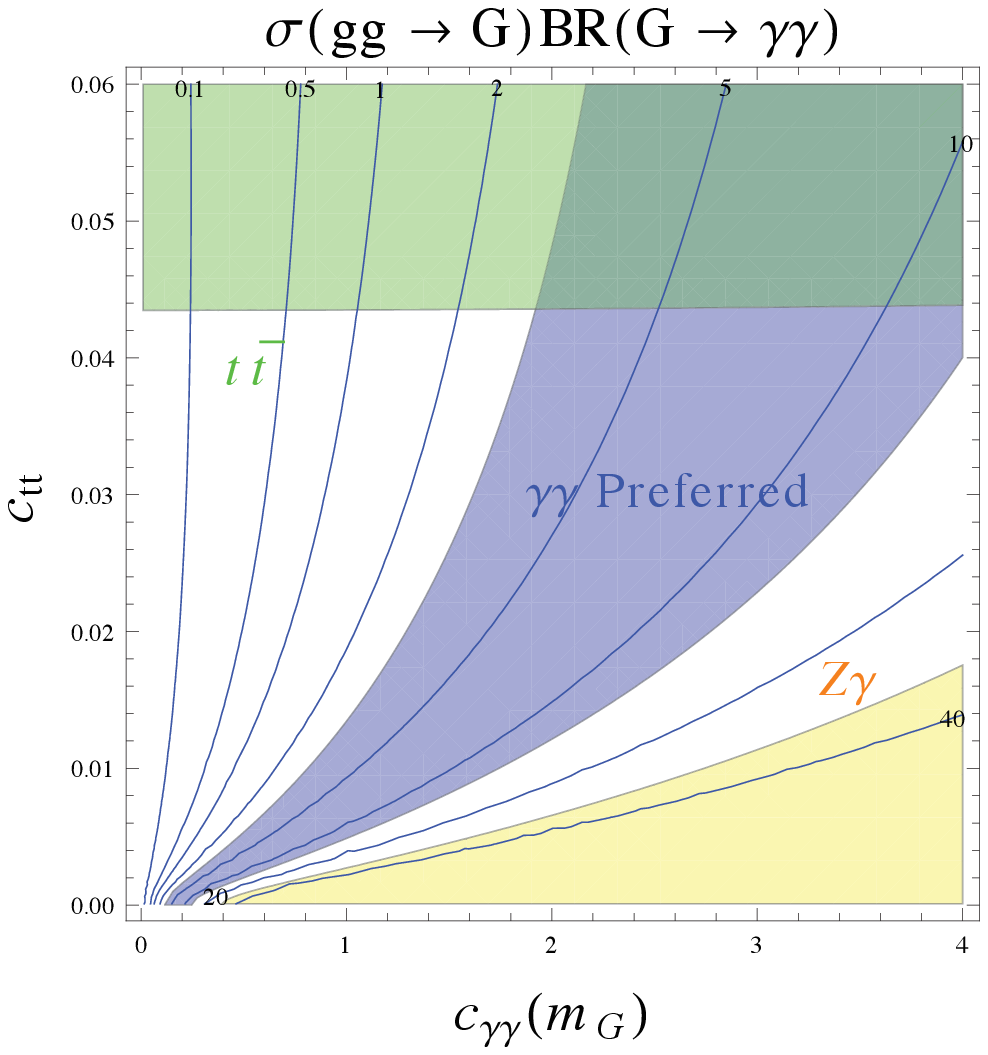}
\caption{Parameter space of the spin-2 particle favored by the $\gamma\gamma$ excess observed by ATLAS and CMS (Blue Region) in the $c_{\gamma\gamma}(m_G)$-$c_{gg}(m_G)$ plane with $c_{tt} = 0.04$ (Left Plot) and $c_{\gamma\gamma}(m_G)$-$c_{tt}$ plane with $c_{gg} (m_G) = 0.03$ (Right Plot), together with the constraints from $t\bar{t}$ (Green Region) and $Z\gamma$ (Yellow Region), where the numbers on the blue lines denote the corresponding $\gamma\gamma$ production cross sections due to spin-2 particle decays.}
\label{pp1}
\end{figure}
In Fig.~\ref{pp1}, we identify the viable parameter space of the spin-2 particle which could explain the ATLAS and CMS $\gamma\gamma$ excesses with the top-quark loop open. The essential constraint only comes from the $t\bar{t}$ channel which limits 
 $c_{gg}(m_G)$ and $c_{tt}$ to small values of ${\cal O}(10^{-2})$. However, in order to explain the $\gamma\gamma$ signal, it requires that $c_{\gamma\gamma}(m_G)$ should be large enough, typically of ${\cal O}(1)$, which implies that the decay of $G\to\gamma\gamma$ goes primarily via this tree-level coupling. In contrast, the diphoton signal still allows the interesting region in which $c_{gg}(m_G)$ can be larger than or of the same order as $c_{tt}$, and thus the top-quark loop plays an
 important role in the ggF production of $G$.

\section{Effects of Invisible Decays of the Spin-2 particle}
As mentioned in the Introduction, the ATLAS data favors a wide resonance of width 45~GeV with the local significance of 3.9~$\sigma$, while the CMS one
shows that the best fit indicates a narrow resonance. However, for a generic spin-2 particle with reasonable couplings, the total decay width from the SM channels is only about of ${\cal O}(1~{\rm GeV})$. Therefore, if the broad width of the resonance is confirmed in the future LHC Run-2 experiments, there should be a large branching ratio for $G$ to decay to some invisible channels, probably related to dark matter. In this section, we consider the effects of this invisible decay to our spin-2 particle phenomenology. Rather than specifying a concrete model, we just parametrize the invisible decay by its partial decay width in our phenomenological discussion.

\begin{figure}[th]
\includegraphics[scale=0.7]{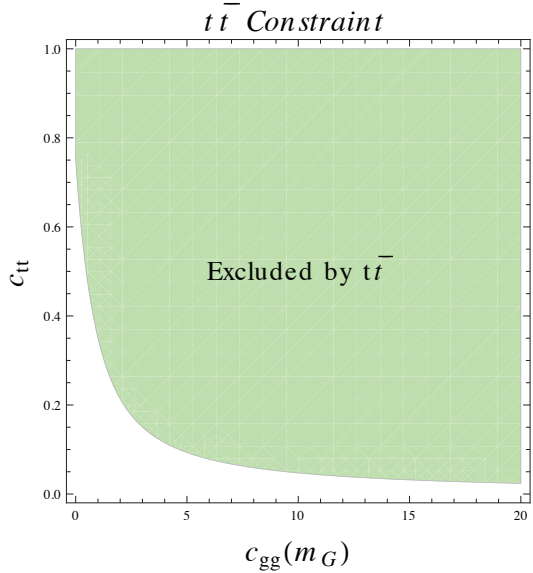}
\caption{The constraint from $gg\to G \to t \bar{t}$ in the $c_{gg}(m_G)$-$c_{tt}$ plane with $c_{\gamma\gamma} (m_G) = 0.1$ and $\Gamma(\rm Invisible) = 40$~GeV, where the green region is excluded by the CMS data~\cite{Khachatryan:2015sma}.}
\label{Ctt_I}
\end{figure}

\begin{figure}[th]
\includegraphics[scale=0.7]{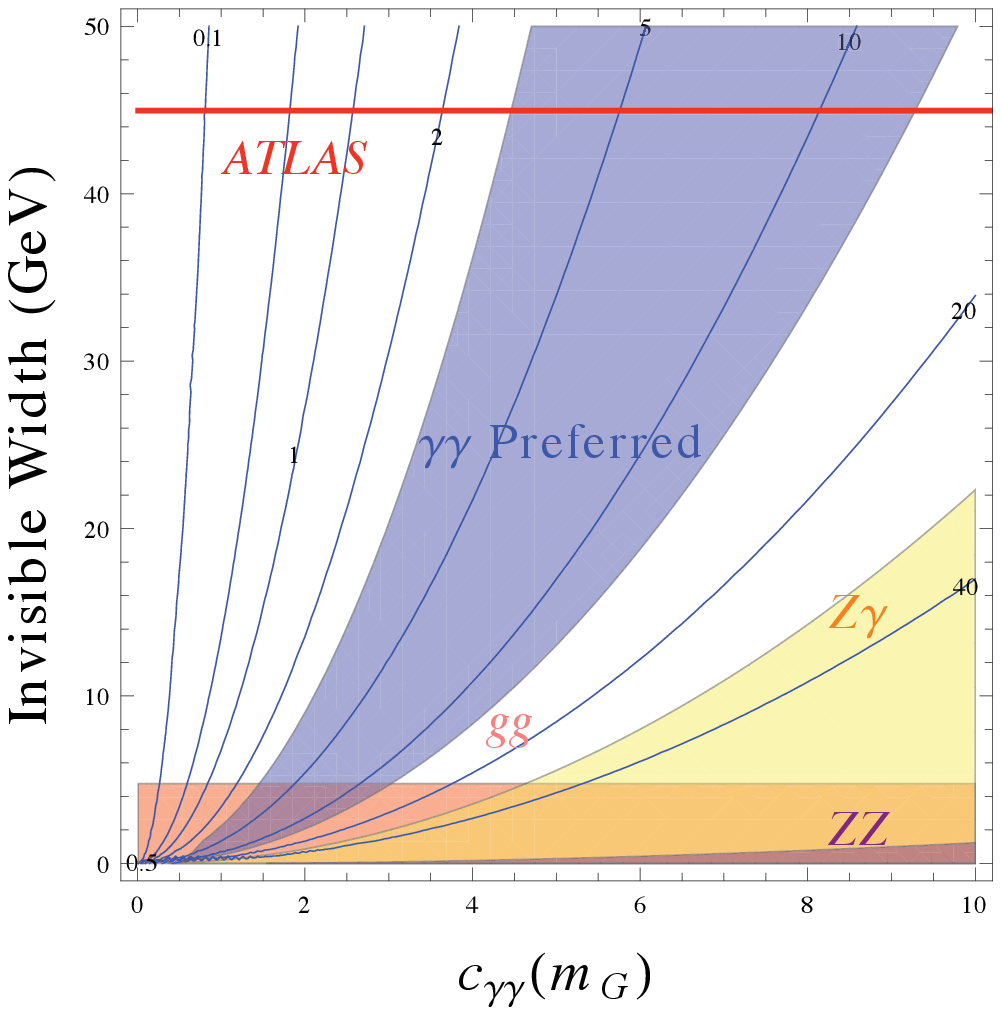}\,\,\,
\includegraphics[scale=0.7]{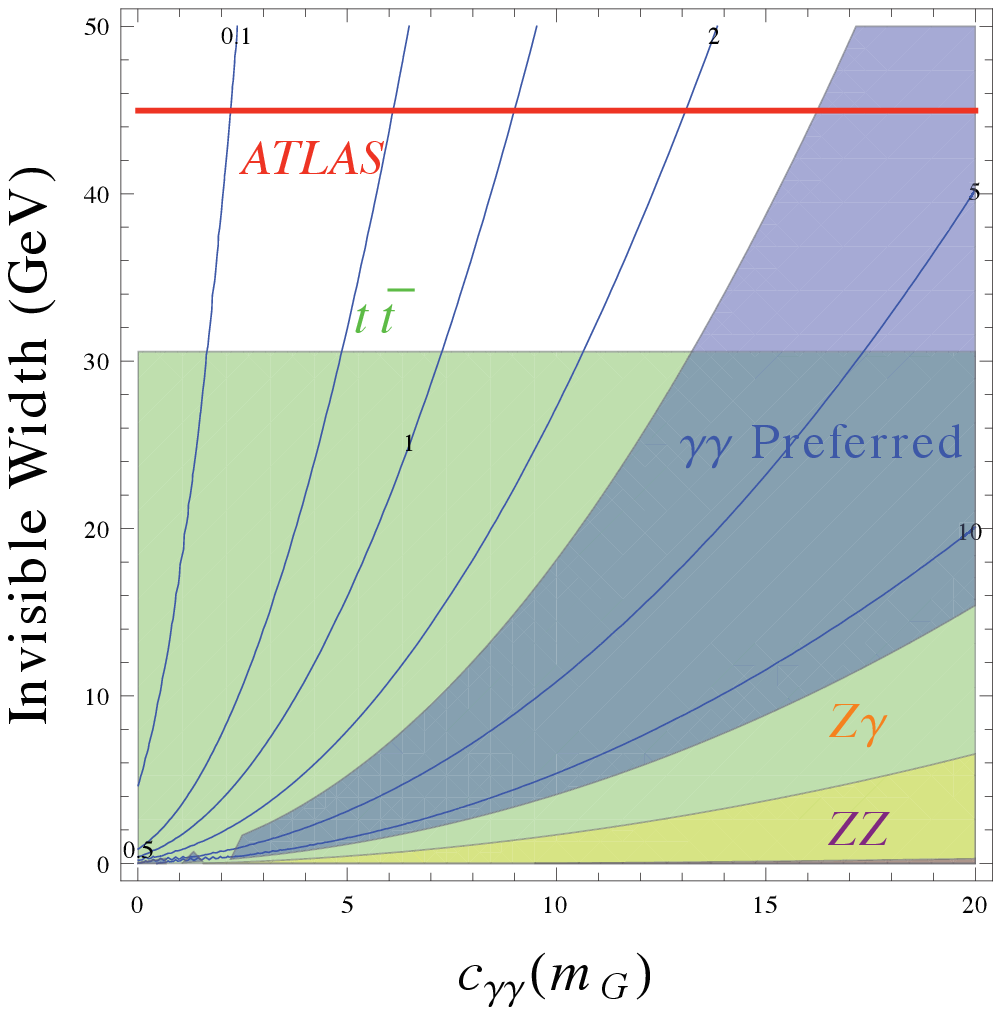}
\caption{Parameter space of the 750~GeV spin-2 particle favored by the $\gamma\gamma$ excess observed by ATLAS and CMS (Blue Region) in the $c_{\gamma\gamma}(m_G)$-$\Gamma_{\rm Invisible}$ plane with $c_{tt} = 0$ and $c_{gg} (m_G) = 5$ (Left Plot) and with $c_{gg} (m_G) = 1$ and $c_{tt} = 0.3$ (Right Plot), together with regions excluded by the constraints from $t\bar{t}$ (Green Region), $Z\gamma$ (Yellow Region), $ZZ$ (Purple Region) and $gg$ (Pink Region), where the numbers on the blue lines denote the corresponding $\gamma\gamma$ production cross sections due to spin-2 particle decays, while
the red lines in both plots denote the total width $\Gamma_{\rm tot} = 45$~GeV favored by the ATLAS data.}
\label{ppI}
\end{figure}

As was done early for the case without invisible decays, we first consider the $t\bar{t}$ constraint, which is shown in Fig.~\ref{Ctt_I}. 
Compared with Fig.~\ref{Const_tt}, the invisible width makes the viable parameter space extend to large values, since it suppresses the $t\bar{t}$ branching ratio greatly. In the figure, the left plot corresponds to the case with the top-quark loop turned off $c_{tt} = 0$,
 while the right one to the case with large top-quark loop contributions $c_{tt} \neq 0$.
Fig.~\ref{ppI} presents the $\gamma\gamma$ decay widths as functions of $c_{\gamma\gamma} (m_G)$ and the invisible decay width $\Gamma_{\rm Invisible}$,
respectively.
It is seen that we can still find parameter regions with large coupling $c_{\gamma\gamma} (m_G)$ that can explain the $\gamma\gamma$ excess, even in the face of the dilution from the invisible decay. It results in that the $\gamma\gamma$ decay is mainly through the tree-level coupling. However, when $c_{tt}$ is nonzero, the top-quark loop can give non-negligible contributions to the spin-2 particle $G$ production via the ggF process. Note also that, in most regions favored by the $\gamma\gamma$ excess, the invisible decay dominates the branching ratios of $G$, which is expected since visible channels cannot fully account for the large width $\sim 45$~GeV implied by the ATLAS data.

\section{Conclusions and Discussions}\label{Summary}
In this paper, we have explored the possibility of a massive spin-2 particle $G$ to interpret the $\gamma\gamma$ excess at 750~GeV recently observed by ATLAS and CMS. In particular, we have focused on the parameter space in which the top-quark loop to diphoton and digluon amplitudes can have the similar order as the corresponding tree-level couplings. Such a scenario can be naturally realized in the generalized warped extra dimension models, in which the gauge 
 and massive KK graviton fields propagate in the bulk, while the top-quark field is localized on the IR brane. By requiring the diphoton production cross section to fit the signal while satisfying the LHC Run-1 constraints of other channels, we have found
  that the preferred parameter space can be separated into two regions with either a nearly vanishing top-quark coupling of $G$ or a relatively large one. It is interesting to note that in the large coupling region, the top-quark loop can compete to and even overwhelm over the tree-level contribution in the $G$ production via the gluon-gluon fusion. We have also considered the case in which the massive spin-2 particle have a large invisible decay branching ratio with the similar results obtained.

Having shown that the massive spin-2 particle is a candidate to explain to the diphoton excess, the next step is to test the present scenario. One immediate question is how to distinguish the spin-2 particle candidate with the (pseudo-)scalar one. As pointed in Refs.~\cite{Han:2015cty,Martini:2016ahj}, the kinematic distributions of the diphoton final states, such as the photon rapidity distribution and the photon angle distribution respect to the beam axis, can provide us with such information. Moreover, for the case with the top-quark loop dominated $G$ production, the largest decay mode would be the $t\bar{t}$ channel, which can be used to distinguish from the case with $c_{tt} = 0$. 
In terms of Figs.~\ref{ProdG} and \ref{BR1}, the $t\bar{t}$ production cross section from the $pp$ collision 
via the 750 GeV massive spin-2 particle is 2~pb, which, together with the total integrated luminosity of ${\cal O}(100)$~fb$^{-1}$ 
at the end of the LHC Run-2, would provide a clear signal to probe the present model. Another kind of the signal for the case with 
the large spin-2 particle width is the invisible decay final states, which are assumed to be the dark matter candidate. 
Since the direct measurement of the total width is limited by statistics, it is helpful to measure some associated channels with 
a large missing transverse energy $E_T$ at the LHC, such as the mono-jet, mono-photon and mono-$Z$. For simplified dark matter models, 
some of the studies have already been done in Ref.~\cite{Han:2015cty}, in which the measurement of the mono-jet plus a missing $E_T$ 
at the LHC~\cite{Aad:2014aqa}  places a tight constraint on the allowed parameter space. 
Nevertheless, we emphasize that the results depend on the details of the dark matter properties, such as the dark matter spin, mass and couplings to SM particles. 
Clearly, a specific and complete construction of the dark sector, associated with the 750 GeV massive spin-2 particle,
 is interesting to be investigated  in the future.

\noindent
\begin{acknowledgments}
The work was supported in part by National Center for Theoretical Science, National Science
Council (NSC-101-2112-M-007-006-MY3), MoST (MoST-104-2112-M-007-003-MY3) and National Tsing Hua
University (104N2724E1).
\end{acknowledgments}

\appendix
\section{Production and Decays of the Massive Graviton}\label{App1}
In this section, we present some details of the calculation of the top-quark loop contributions to the massive spin-2 particle $G$ couplings to the diphoton and digluon, as shown in Fig~\ref{FeynD}. For concreteness, we first focus on the $G\gamma\gamma$ effective coupling induced by the top-quark loops. Note that in the model presented in Sec.~\ref{Model} only the right-handed top quark couples to spin-2 particle, so that we apply the Feynman rules listed in Ref.~\cite{Han:1998sg} with the right-handed chiral projection in the vertices for top quarks. The explicit calculation of Feynman diagrams in Fig.~\ref{FeynD} gives the the amplitude for the top-quark loop induced $G$ decay as
\begin{eqnarray}\label{amp}
-i{\cal M}_1(G\to \gamma\gamma) &=& \epsilon_\rho^{A *} (k_1) \epsilon_\sigma^{B *} (k_2) \epsilon^s_{\mu\nu}(P) [C_{\mu\nu,\rho\sigma}k_1\cdot k_2 + D_{\mu\nu,\rho\sigma}(k_1, k_2)]\nonumber\\
&&  (-i)\frac{\alpha Q_t^2 N_c}{6\pi} \frac{c_{tt}}{\Lambda} [\frac{2}{\epsilon}-\gamma + \ln\frac{4\pi\mu^2}{m_G^2} + A_G\left(\frac{4 m_t^2}{m_G^2}\right)]\,,
\end{eqnarray}
with the finite part of the integral $A_G(\tau)$ defined as
\begin{eqnarray}\label{ag}
A_G(\tau) &=& -\frac{1}{12}\Big[ -\frac{9}{4} \tau (\tau + 2) [2{\rm tanh}^{-1}(\sqrt{1-\tau}) - i \pi]^2 + 3 (5\tau + 4) \sqrt{1-\tau}[2{\rm tanh}^{-1}(\sqrt{1-\tau})-i\pi] \nonumber\\
& &  -39 \tau + 12\ln \tau -35 - 12\ln4  \Big]\,,
\end{eqnarray}
where $Q_t$, $m_t$ and $N_c$ denotes the charge, mass and color factor of the top quark. The tensor factors $C_{\mu\nu, \rho \sigma}$ and $D_{\mu\nu, \rho\sigma}$ are defined as in Appendix of Ref.~\cite{Han:1998sg}.  Note that the formula for $A_G(\tau)$ in Eq.~(\ref{ag}) is only valid when $\tau < 1$, which is the case when $m_G = 750$~GeV.  In addition, in Eq.~(\ref{amp}), we keep the terms up the order of ${\cal O}(k^2)$ while ignoring the higher-order ones. Concretely, when deriving Eq.~(\ref{amp}), we have encountered a term of ${\cal O}(k^4)$:
\begin{equation}\label{T4}
C k_{1\mu} k_{2\nu} k_{2\rho} k_{1\sigma}\,,
\end{equation} 
where $C$ is a constant dimensionless parameter. This term can induce a dimension-7 effective operator like $G_{\mu\nu} \partial^\mu F^{\rho\sigma} \partial^\nu F_{\rho\sigma}$, which, from the usual effective field theory perspective, is expected to be suppressed by higher powers of cutoffs or $m_G^2$ so that it can be negligible. However, there is a subtlety to do this. Since Eq.~(\ref{T4}) is not gauge covariant under the electromagnetic $U(1)$ symmetry, we need to first transform it to the following form
\begin{eqnarray}
C k_{1\mu} k_{2\nu} ( k_{2\rho} k_{1\sigma} - \eta_{\rho \sigma} k_1 \cdot k_2) + C k_{1\mu} k_{2\nu} \eta_{\rho \sigma} k_1 \cdot k_2\,.
\end{eqnarray}
In this way, the first term in the parentheses is now gauge covariant and can be safely neglected, while the second one gives the extra contribution
\begin{eqnarray}
C (m_G^2/2) \eta_{\rho \sigma} k_{1\mu} k_{2\nu}\,,
\end{eqnarray}
with the on-shell relation $k_1 \cdot k_2 = m_G^2/2$. Only with this method we can obtain the correct Lorentz structure for the massive graviton-photon vertex in Eq.~(\ref{amp}). 

If we consider the tree-level contribution from the direct massive graviton-photon vertex in Eq.~(\ref{LagG}), we have
\begin{eqnarray}
-i{\cal M}_0 (G\to \gamma\gamma)  =  \epsilon_\rho^{A *} (k_1) \epsilon_\sigma^{B *} (k_2) \epsilon^s_{\mu\nu}(P) [C_{\mu\nu,\rho\sigma}k_1\cdot k_2 + D_{\mu\nu,\rho\sigma}(k_1, k_2)] \left(-i \alpha c^0_{\gamma\gamma}/(4\pi \Lambda)\right),\nonumber \\
\end{eqnarray}
where $C_{\gamma\gamma}^0$ represents the bare couplings defined in the Lagrangian, which should be renormalized to obtain a sensible result.

The total amplitude for the decay of $G$ into two photons is
\begin{eqnarray}
-i{\cal M} (G\to \gamma\gamma)  &=&  \epsilon_\rho^{A *} (k_1) \epsilon_\sigma^{B *} (k_2) \epsilon^s_{\mu\nu}(P) [C_{\mu\nu,\rho\sigma}k_1\cdot k_2 + D_{\mu\nu,\rho\sigma}(k_1, k_2)] \nonumber\\
&& (-\frac{i\alpha}{4\pi \Lambda})\left[c_{\gamma\gamma}^0 + \frac{2}{3} c_{tt} Q_t^2 N_c \left(\frac{2}{\epsilon}-\gamma + \ln\frac{4\pi\mu^2}{m_G^2} + A_G(\frac{4 m_t^2}{m_G^2})\right)\right]\,.
\end{eqnarray}
If we renormalize the Wilson coefficient as
\begin{eqnarray}
c_{\gamma\gamma}(\mu) = c_{\gamma\gamma}^0 + \frac{2}{3} c_{tt} Q_t^2 N_c \left(\frac{2}{\epsilon} - \gamma + \ln (4\pi)\right)\,,
\end{eqnarray}
then the renormalized amplitude is
\begin{eqnarray}
-i{\cal M} (G\to \gamma\gamma)  &=&  \epsilon_\rho^{A *} (k_1) \epsilon_\sigma^{B *} (k_2) \epsilon^s_{\mu\nu}(P) [C_{\mu\nu,\rho\sigma}k_1\cdot k_2 + D_{\mu\nu,\rho\sigma}(k_1, k_2)] \nonumber\\
&& (-\frac{i\alpha}{4\pi\Lambda})\left[c_{\gamma\gamma} (\mu) + \frac{2}{3} c_{tt} Q_t^2 N_c \left( \ln\frac{\mu^2}{m_G^2} + A_G(\frac{4 m_t^2}{m_G^2})\right)\right]\,.
\end{eqnarray}
If the renormalization scale is chosen to be $\mu = m_G$, then
\begin{eqnarray}
-i{\cal M} (G\to \gamma\gamma)  &=&  \epsilon_\rho^{A *} (k_1) \epsilon_\sigma^{B *} (k_2) \epsilon^s_{\mu\nu}(P) [C_{\mu\nu,\rho\sigma}k_1\cdot k_2 + D_{\mu\nu,\rho\sigma}(k_1, k_2)] \nonumber\\
&& (-\frac{i\alpha}{4\pi\Lambda})\left[c_{\gamma\gamma} (m_G) + \frac{2}{3} c_{tt} Q_t^2 N_c  A_G\left(\frac{4 m_t^2}{m_G^2}\right)\right] \,.
\end{eqnarray}
Now we define the effective $G\gamma\gamma$ coupling as
\begin{eqnarray}\label{cEeff}
c_{\gamma\gamma}^{\rm eff} = c_{\gamma\gamma} (m_G) + \frac{2}{3} c_{tt} Q_t^2 N_c  A_G\left(\frac{4 m_t^2}{m_G^2}\right) \, ,
\end{eqnarray}
so that the partial decay width for the diphoton channel is
\begin{eqnarray}
\Gamma (G \to \gamma\gamma) = \frac{m_G^3}{80\pi\Lambda^2} \left|\frac{\alpha c_{\gamma\gamma}^{\rm eff}}{4\pi} \right|^2 \,.
\end{eqnarray}

For the digluon channel, we have the similar results except that the charge factor in Eq.~(\ref{cEeff}) should be replaced by the corresponding color factors
\begin{equation}
c_{gg}^{\rm eff} = c_{gg} (m_G) + \frac{c_{tt}}{3}  A_G\left(\frac{4 m_t^2}{m_G^2}\right)\,.
\end{equation}
If we further consider the fact that gluons are in the color adjoint representation, the decay can be easily obtained
\begin{eqnarray}
\Gamma (G \to gg) = \frac{m_G^3}{10\pi\Lambda^2} \left|\frac{\alpha c_{gg}^{\rm eff}}{4\pi} \right|^2 \,.
\end{eqnarray}

For the $Z\gamma$ and $ZZ$ channels we can obtain the similar expressions by calculating the corresponding Feynman diagrams as in Fig~\ref{FeynD}. It is interesting to note that due to our assumption that only the hypercharge $U(1)_Y$ gauge bosons couple to the massive spin-2 particle and the right-handed top-quarks, the renormalization of these two processes do not involve new counter terms so that the renormalization in the diphoton channel is enough to make the amplitudes of $Z\gamma$ and $ZZ$ finite, as shown in Eqs.~(\ref{ZpRate}) and (\ref{ZZRate}).

\end{document}